\documentclass[10pt]{article} 
\usepackage[a4paper,margin=0.75in]{geometry}

\usepackage{times} 
\usepackage[hyphens]{url}  
\usepackage{graphicx}
\usepackage{titlesec}

\titlespacing*{\section}
  {0pt}     
  {0.5em}   
  {0.5em}   

\titlespacing*{\subsection}
  {0pt}
  {0.5em}
  {0em}

\usepackage[authoryear]{natbib} 
\bibpunct{(}{)}{;}{a}{,}{,}
\usepackage[dvipsnames]{xcolor}
\usepackage[pagebackref, colorlinks=true, linkcolor=MidnightBlue, citecolor=MidnightBlue, urlcolor=MidnightBlue]{hyperref}
\usepackage[english]{babel} 
\renewcommand*{\backref}[1]{}
\renewcommand*{\backrefalt}[4]{%
  \ifcase #1 %
    Not cited.%
  \or
    Cited on page~#2.%
  \else
    Cited on pages~#2.%
  \fi}

\usepackage{caption}
\usepackage[most]{tcolorbox}
\usepackage{enumitem}
\title{A Moral Agency Framework for Legitimate Integration of AI in Bureaucracies}
\usepackage{authblk} 
\author[1]{Chris Schmitz}
\author[1]{Joanna Bryson}
\affil[1]{Centre for Digital Governance, Hertie School, Berlin}
\date{}

\begin{document}

\twocolumn[
  \begin{@twocolumnfalse}
    \maketitle
    \begin{abstract}
      Public-sector bureaucracies seek to reap the benefits of artificial intelligence (AI), but face important concerns about  accountability and transparency when using AI systems. In particular, perception or actuality of AI agency might create ethics sinks --- constructs that facilitate dissipation of responsibility when AI systems of disputed moral status interface with bureaucratic structures. Here, we reject the notion that ethics sinks are a necessary consequence of introducing AI systems into bureaucracies. Rather, where they appear, they are the product of structural \textit{design decisions} across both the technology and the institution deploying it. We support this claim via a systematic application of conceptions of moral agency in AI ethics to Weberian bureaucracy. We establish that it is both desirable and feasible to render AI systems as tools for the generation of organizational transparency and legibility, which continue the processes of Weberian rationalization initiated by previous waves of digitalization. We present a three-point Moral Agency Framework for legitimate integration of AI in bureaucratic structures: (a) maintain clear and just human lines of accountability, (b) ensure humans whose work is augmented by AI systems can verify the systems are functioning correctly, and (c) introduce AI only where it doesn't inhibit the capacity of bureaucracies towards either of their twin aims of legitimacy and stewardship. We suggest that AI introduced within this framework can not only improve efficiency and productivity while avoiding ethics sinks, but also improve the transparency and even the legitimacy of a bureaucracy.
    \end{abstract}
    \vspace{1em}
  \end{@twocolumnfalse}
]

\section{Introduction}

Public sector organizations (PSOs) are increasingly integrating artificial intelligence (AI) systems into their operations, hoping to capture the gains in efficiency, personalization, and worker satisfaction the technology promises. Existing research identifies a wide range of tasks within PSOs  potentially addressable  using AI technology, including automating routine processes, interacting with citizens, structuring unstructured data, and supporting decision-making \citep{wirtz_2019_ArtificialIntelligencePublicSectorApplications, hjaltalin_strategic_2024}. Significant capability gains for AI tools have fueled predictions that this range will continue to expand and may soon include the vast majority of tasks currently performed by humans in PSOs. Further, the ability of non-state actors to leverage these capability gains may put PSOs under an imperative to integrate AI systems into their operations to retain their relative capacity  \citep{bullock_2025_AGIGovernmentsFreeSocieties}.

Ideally, the introduction of AI systems should simply continue processes of rationalization that previous waves of digitalization have driven \citep{muellerleile_digital_2018,mokander_2024_ArtificialIntelligenceRationalizationlimitscontrol}. Such processes include the generation of organizational transparency, legibility, and legitimacy via encoding processes in a formal logic \citep{doi:10.1177/09520767231197801}, and retaining complete and searchable digital records of processes, decisions, and accountability.

However, at least two narratives of AI technology run counter to this outcome. First, AI is often characterized as `necessarily opaque,' ignoring a vast literature and indeed even established standards on AI transparency \citep{winfield_ieee_2021}. Second, there is debate on whether AI systems can hold responsibility for their actions, a property known as \textit{moral} \textit{agency}. Appropriate attribution of moral or at least legal agency within any organization is critically important. Bureaucracies in particular derive their democratic legitimacy from their ability to correctly attribute responsibility and ensure accountability, in order to ensure not only good behavior, but steady improvement \citep{gasser_guardrails_2024}. 

Thus both disputed moral status and more general claims of intransparency have led to concerns about introducing opaqueness, accountability gaps, or an entirely new and largely unaccountable form of agency into bureaucracies. Such concerns can even be self-fulfilling. Nothing about digital systems intrinsically ensures that they produce the records necessary for accountability, or that such records, if produced, are stored immutably. Thus myths of unaccountability could lead to the procurement of AI systems unsuitable for good government. 

Here, we seek to refute the notion that unresolvable questions of moral agency and accountability are an inevitable consequence of introducing AI into bureaucratic structures. Rather, constructs leading to unattributed accountability in bureaucracies --- which we term as “ethics sinks” --- are the result of \textit{design decisions}. It is feasible both institutionally and technically to integrate AI systems within robust and holistic systems of responsibility distribution, and thereby to ensure that AI systems enhance, rather than inhibit, bureaucratic rationality. We postulate that there exist frameworks of design rules, which allow organizations to  avoid the misapplication of power, and to ensure the accountability essential for social maintenance and improvement. Indeed, we propose one such framework here. In so doing, we seek to render any institutional design that does afford ethics sinks as evidently negligent.

This paper is structured as follows. In Section~\ref{dual.sec}, we review from established literature the structural promise and purpose of bureaucracy, as relevant for the context of the responsible introduction of AI. We describe the role of moral agency in enabling the dual aims of bureaucracy: legitimate enactment of legislation, and the provision of stewardship that allows social maintenance and improvement. In Section~\ref{sit.sec}, we review the debate as to whether AI is a tool or a moral subject. We dismantle this false dichotomy in a bureaucratic context by framing AI systems as elements of the \textit{institutional structure} of Weberian bureaucracy. In systems designed with care, machine intelligence supports the moral agency of individual bureaucrats, without siphoning it off to other organizations, or itself holding moral status. This allows bureaucracies to be grounded in democratic legitimacy, provided that all (human) moral agents are answerable to elected officials.  In Section~\ref{fw.sec}, we propose a three-step moral agency framework (MAF) for integrating AI into bureaucracies. 
We show that the framework prevents issues of responsibility attribution arising from the use of ``ordinary" agents --- active entities holding no moral status --- by the human network that composes the moral core of a bureaucracy.

\section{The Dual Purposes of Bureaucracy} \label{dual.sec}
Bureaucratic public-sector organizations serve twin purposes. The first is the legitimate enactment, implementation, and enforcement of legislation passed down by the elected legislature. The other is the provision of stewardship, that is: the maintenance of conditions of long-term stability for both the general public, and the government itself. In this section, we establish that the mechanisms by which bureaucracy fulfills each of these purposes presuppose, and even require, the moral agency of human bureaucrats. We further argue that to the extent a trade-off exists between bureaucracy's twin purposes, the introduction of AI systems --- or any other measures impacting the agency of the organization --- can aid one purpose while interfering with the other. We argue such a perverse outcome is plausible, but may not be necessary, and should be defended against.

\begin{figure}[t]
    \centering
    \includegraphics[width=0.9\linewidth]{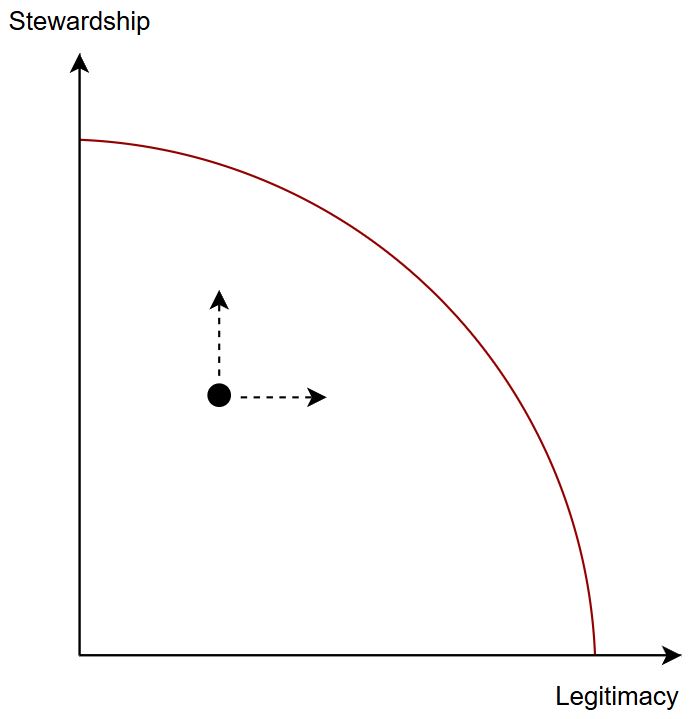}
    \captionof{figure}{Illustration of the ``Pareto Frontier" between bureaucracies' dual aims of stewardship and legitimacy. PSOs not at the frontier can improve in one dimension without necessarily jeopardizing the other.}
    \label{fig:pareto}
\end{figure}

\subsection{Legitimate Implementation of Legislation}

In most government systems, public sector organizations (PSOs) have a considerable mandate to interpret, enforce, and uphold legislative will. They are usually granted power to take actions with significant and wide-ranging implications for citizens. This power is, in democracies, also granted by these citizens, as it is tied to the promise of fulfilling legislative goals set by elected representatives. As \citet{habermas_between_2015} argues, PSOs in democracies are therefore only legitimate to the extent to which the public can trust them to work wholly, reliably and uncompromisingly towards these legislative goals. Under what \citet{weber1978economy} describes as the ``compliance model'' of bureaucracy, a core purpose of bureaucracy as an organizational structure is to enable this trust. 

The design of Weberian bureaucracy serves this goal: to make organizations legitimate by making them legible and faithful implementers of legislation. 
The resultant organizational transparency is what \citet{weber1978economy} 
refers to as bureaucratic rationality. Weber famously stresses the machine-like qualities of ideal-type bureaucracy: “The fully developed bureaucratic apparatus compares with other organizations exactly as does the machine with the non-mechanical modes of production”. 
In this characterization, bureaucracy’s structural features --- hierarchy, delegation, specification, and impersonal procedure, to name a few --- serve simultaneously to make the organization legible and to ensure it is working precisely towards specified goals. The organization is an ``intermediary" between legislative inputs and real-life outcomes \citep{latour_reassembling_2005, zacka_2022_PoliticalTheoryRediscoversPublicAdministration} --- in the best case, it is invisible. 

Weber’s ideal-type model has been refined and challenged frequently, notably by \citet{simon_administrative_1947}, who affords more room for unspecified and adaptive processes within bureaucratic bodies. Further, in most democracies these features of the institutions themselves are reinforced by extra-institutional checks and systems, notably strong traditions of administrative law, administrative court systems, and freedom of information regulation \citep{pozen_transparencys_2018}. Nonetheless the motivations of Weber’s model, as well as the practical capacity of these structural features to meet those motivations, have held up remarkably well. As \citet{newman_2022_DigitalTechnologiesArtificialintelligencebureaucratic} put it, “bureaucracy has proven remarkably resilient.''

Human adults such as bureaucrats are by nature \emph{moral agents}; that is, they are expected to hold responsibility for their actions. Bureaucrats have considerable space and range of \emph{agency} --- they can considerably alter their world through actions (including inactions), by activating policies and procedures.  Bureaucrats also have individual opinions and moral convictions. A core achievement of bureaucracies is reconciling this plurality with the rationality of entire organizations. Individual bureaucrats facilitate this by entering Weber’s “iron cage” --- a construct in which they are expected to become more cog-like, readable, and predictable \citep{clegg_2009_WeberSinteringIronCageTranslation}. Weber conceives of the ideal civil servant as executing orders of the superior authority ``exactly as if the order agreed with their own convictions" \citep{shafir_incongruity_1985}. Their individual agency is thus constricted, and their personal morality subordinated to their professional morality and their oath to their office. In exchange, they carry little or no blame for decisions they make, and actions they take, in their professional role --- so long as these are compliant with the organizational restrictions they are under, including transparent accountability to the legitimate, root moral actors. 

\citet{hood_2006_PoliticsPublicServiceBargainsReward} expand this agency trade-off into the ``public service bargain”: a moral-economical deal bureaucrats enter which reduces variance across many dimensions of their life.  In exchange, they receive stability --- fixed job progression in exchange for higher job security, fixed areas of responsibility in exchange for lower expectation of on-job growth, and restrictions of agency and discretion in exchange for diffused responsibility for their actions. These deals are generally attractive, \citeauthor{hood_2006_PoliticsPublicServiceBargainsReward} argue. The continued popularity of public service work supports their claim.

Bureaucracies require not only the (somewhat-homogenized) moral agency of bureaucrats for these deals to materialize, but also their basic humanity, because these trade-offs rely on mechanisms of discipline and control. To enable institutional rationality, individual bureaucrats must be incentivized to act according to their professional morality, and disincentivized from instead following their personal interests --- or convictions. As \citet[p. 369]{gailmard_2012_FormalModelsBureaucracy} put it, “Accountability in the standard principal–agent setting requires rewarding or punishing the agent for the degree to which his or her choice of action is consistent with what the principal would have him or her do in the realized state.” Participation in structures of responsibility attribution --- justice systems, to generalize --- therefore presupposes a responsiveness to the means these systems have to reward “good” and punish “bad” behavior. Justice, in short, requires accountability \citep{bryson_2017_PeopleLegalLacunasyntheticpersons, gasser_guardrails_2024}. Any entity that is to participate in these moral structures must  be responsive to the persuasions and dissuasions of the law. The law requires a guaranteed responsiveness to penalty. 

Human responsiveness to penalty is derived in part from our ability to experience conscious suffering \citep{bentham_2005_IntroductionPrinciplesMoralsLegislation}. We are by nature exposed to what \cite[p. 4]{sebo_2018_MoralProblemOtherMinds} refers to as “a private, subjective, qualitative experience.”  Humans are responsive to judicial incentives and disincentives not just because we feel physical pain, but because we have rich, conscious (actionable) positive and negative reactions, which judicial systems can harness to ensure compliance with their laws. For example, we value the time in our limited lifespans, and have a pervasive aversion to spending parts of it in the limitations of jail. Similarly, we value maintaining or increasing our job rank. We appreciate promotions, in large part (but not entirely) because these have implications for the economic security of ourselves and our families. We also tend to enjoy power and discretion.

This is what enables the mechanisms of ``discipline and control" which constrain individual bureaucrats \citep{weber1978economy}. These are also the quale ordinarily seen as qualifying us for moral patiency --- the duty of care society's actors have towards us \citep{floridi_2004_MoralityArtificialAgents}. But to be clear, bureaucracies require not so much the moral patiency of bureaucrats itself, but rather their responsiveness to penalty. Nevertheless, this responsiveness rests on characteristics often seen as preconditions qualifying entities to be moral patients \citep{mayerfeld1999suffering}.

\subsection{Stewardship}

Stewardship is defined by the ACSH as ``the perennial mission of the public service; the preservation of the long-term capability of state institutions to act for the greater public good" \citep{acsh_stewardship_2023}. It is increasingly recognized in political philosophy that bureaucratic PSOs are not mere enactors of legislation, but entities which serve a unique structural purpose: the provision of an environment of stability for both the public and other branches of government \citep{zacka_2022_PoliticalTheoryRediscoversPublicAdministration, heath_2020_MachineryGovernmentPublicAdministrationLiberal}.  The public service is attributed this role out of recognition that no other arm of government is positioned to provide this stability: the judiciary is too reactive, the legislative too short-sighted and impulsive.

\citet{heath_2020_MachineryGovernmentPublicAdministrationLiberal} enumerates in detail the advantages this ``rigidity" of the public service has for democracies: ensured day-to-day operation, long-term planning beyond electoral cycles, consistency and correction for ``oversteering" governments, and better state information use \citep[cf.][]{scott_seeing_1998}.  These effects elevate bureaucracies beyond neutral, implementing entities, as Weber's compliance model might be taken to indicate. The effects therefore also reframe commonly-criticized aspects of bureaucratic organizations, such as slowness to respond to legislative change and budget maximization, as --- at least partially --- desirable attributes of a civil service \citep{niskanen_bureaucracy_1971, gay_praise_2000} 

The ability of PSOs to provide stewardship is owed in large part to the moral agency and discretionary space that {\em is} afforded to street-level bureaucrats. \citet{lipsky_street-level_1980} and \citet{simon_administrative_1947} argue that moral agency is required simply for reasons of information capacity: it is infeasible to specify instructions and protocol covering the entire variety of cases bureaucrats are confronted with. The capacity of a bureaucracy to function continuously, in this model, rests on the ability of public servants to interpolate and interpret passed-down instructions --- an ability deriving from moral agency. Examples of such tasks include the specification of vague legislation, the parsing of conflicting instructions, and more generally the countless momentary and ad-hoc decisions bureaucrats make. 

The contribution of individual moral agency to institutional stewardship goes beyond simple discretionary capacity. \citet{zacka_2017_WhenStateMeetsStreetPublic} argues persuasively that it is the \textit{plurality} of moral dispositions individuals have, and use when completing discretionary tasks, that produces the emergent property of stewardship. These moral dispositions across civil servants extend, correct, and complement each other, creating --- within the confines of the bureaucratic structure --- an environment of stability, and enabling actions like whistleblowing and questioning instructions. Unlike that described by \citet{lipsky_street-level_1980}, this form of dependence on individual moral agency cannot be ``rationalized away" by more capable information processing systems --- in fact, it is threatened by them. 

Increasing a bureaucracy's total agency via rationalizing technologies, such as AI, could limit the within-bureaucracy heterogeneity and autonomy on which effective stewardship of the law depends.  \citet{vredenburgh_ai_2023}, for example, argues that AI systems are best characterized by an ``indifferent" archetype of \citeauthor{zacka_2017_WhenStateMeetsStreetPublic}'s street-level bureaucrat. This may be optimistic, since AI systems may in fact be biased towards the interests of (for example) their providers \citep{evans_2023_WeCollaborateWhatWeDesign}. But even if neutral AI is procured,  \citeauthor{vredenburgh_ai_2023} argues a widespread introduction of identically-disposed AI systems jeopardizes the plurality of moral dispositions \citeauthor{zacka_2017_WhenStateMeetsStreetPublic} describes. 

We need to reject any implicit anthropomorphism entailed in a notion of ``identically-disposed systems.'' There are no countable number of (homogenous) AI systems used to replace some (countable) number of humans. It's worse than that.  AI `efficiencies' may indeed reduce both plurality and resilience unless human diversity is valued and maintained, human autonomy is not reduced, and the numbers of employees is sufficient to absorb crises and turnover while maintaining institutional knowledge.  A plausible alternative is a significant weakening of bureaucracies' ability to resist even `legitimate' bouts of policy extremism \citep{Heldring23,bullock_2025_AGIGovernmentsFreeSocieties, barez_toward_2025}.

Further, over-dependencies on single technologies can also lead to a brittleness, fragility, loss of resilience, or simple insecurity by bringing too much value for a government resting on single systems of too little value to their providers. An imbalance here could result in inadequate support or inadequate cybersecurity. The product or service could also simply be terminated if the provider goes bankrupt or loses interest in the product. An AI system a government depends on becomes a potential tool of blackmail or ransom. 

 From this it becomes evident that indeed, in any practical context, a trade-off does exist between the dual aims of stewardship and responsiveness to legislation. As in nature, what must change more quickly must be represented more plastically \citep{hinton1987learning, bryson_2018_PatiencyNotVirtuedesignintelligent}. Structural changes which increase a PSO's capacity to implement legislation fully (and quickly) are likely to achieve this via optimizations of information processing, which may compromise the above-listed benefits of bureaucratic rigidity. Integration of these systems must therefore be informed by a clear understanding of their impact on this balance.

\section{Situating AI in Bureaucracy} \label{sit.sec}

\begin{figure}[bt]
    \centering
    \includegraphics[width=0.8\linewidth]{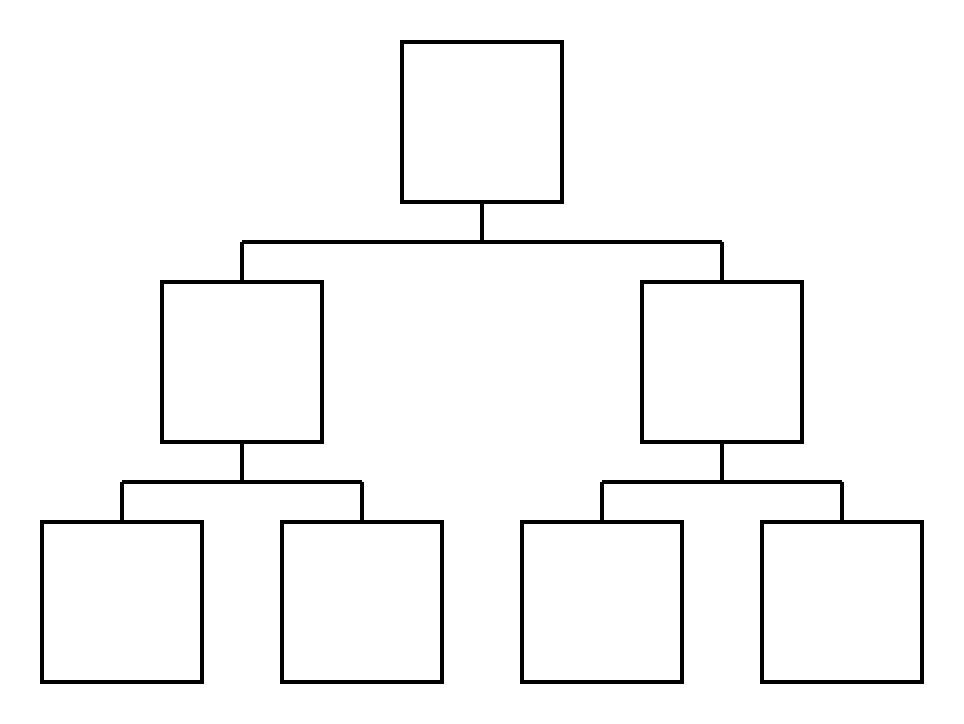}
    \captionof{figure}{The ideal-type Weberian bureaucracy is a system of responsibility attribution between legitimate moral agents. The grounding of that legitimacy in democracies lies in rooting the chains of responsibility in elected representatives.}
    \label{fig:bureaucracy}
\end{figure}

We believe AI systems used in the public sector are best viewed as continuations of Weberian rationalization processes initiated by previous waves of digitalization. Previous work by others has critiqued such suggestions because they require AI systems to be considered ``tools" with no moral impact, rather than as ``more-than-tool" entities worthy of moral consideration. In this section, we first establish that this is a false dichotomy by arguing neither of these accounts fully capture the role of AI systems in bureaucracies. We then argue that it is readily possible, and also desirable, for an AI system to be an artifact with no attributed moral status of its own, which nonetheless bridges and amplifies the moral agency of the responsible --- human --- constituents of a bureaucracy. We posit that this is the same status already afforded to the institutional structure and mechanisms of Weberian bureaucracy, and AI systems are therefore best conceived as such. 

\subsection{AI Systems as Tools}

\begin{figure}[bth]
    \centering
    \includegraphics[width=0.8\columnwidth]{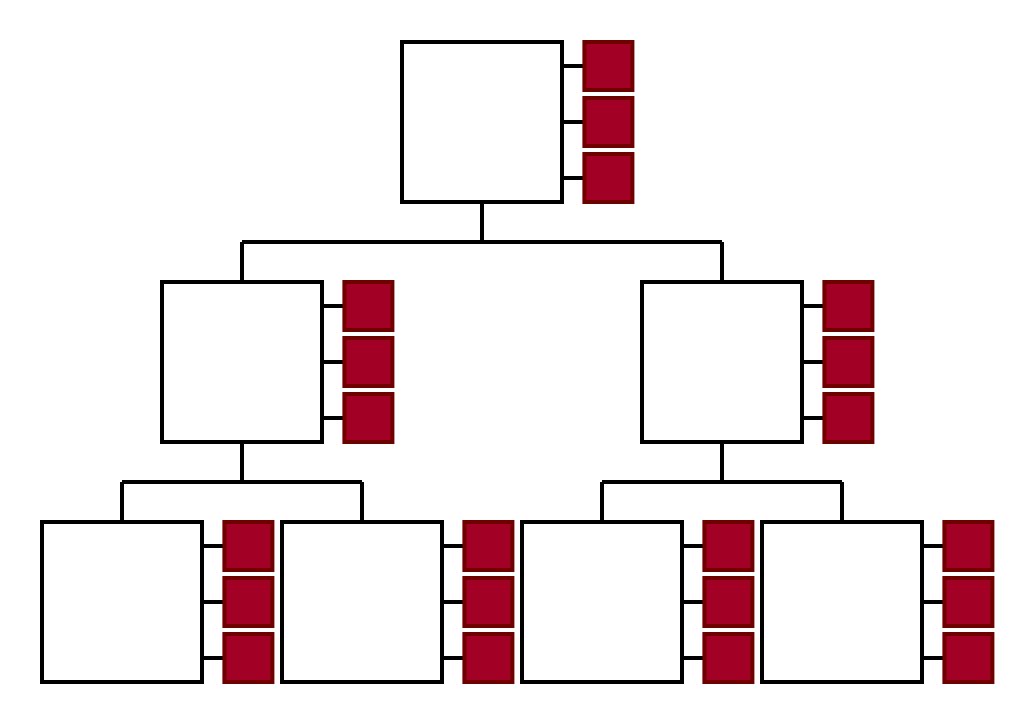}
    \captionof{figure}{Illustration of AI integration in bureaucracy under the ``tool conception". Humans (in white) use AI systems (in red) in purpose-bound ways, thereby retaining moral agency.}
    \label{fig:tools}
\end{figure}

The tool conception of AI systems describes these systems as continuations of previous waves of ICT-driven bureaucracy digitalization and rationalization. \citep{lee_bureaucracies_1984, newman_2022_DigitalTechnologiesArtificialintelligencebureaucratic}. Practical implementation of these systems, which has been purpose-bound, project-based, and sporadic, has so far largely been consistent with this conception \citep{mellouli_introduction_2024}. This is convenient, as tools generally have well-established moral frameworks around them, which keep responsibility solidly in the hands of human agents --- even where tools are ``persuasive" to humans \citep{verbeek_materializing_2006}. The impact that introduction of ICT-based tools has on bureaucratic structures is also maturely theorized, notably in the  \citet{bovens_street-level_2002} model of the system-level bureaucracy. Following this model, ICT systems, which are strictly rule-based and afford record-keeping naturally, can aid the rationalization of bureaucracies. The integration of AI systems, ideally, is a continuation of this rationalization \citep{mokander_2024_ArtificialIntelligenceRationalizationlimitscontrol, bullock_2019_ArtificialIntelligenceDiscretionBureaucracy, bell_replacing_2021}. 

The tool conception faces numerous practical challenges. Firstly, it is questioned by recent developments in the architecture and capabilities of available AI systems. Heidegger's idea of ``\emph{Zuhandenheit}" defines tools as objects that are (a) purpose-bound, and (b) given meaning though human application to problems \citep{heidegger_being_2008, gunkel_other_2018}. Increasingly though AI systems are being primarily seen as the outputs of a small number of providers of global reach. These providers are producing increasingly general-purpose systems, which are increasingly presented as agentic. 

The market presently appears to be moving away from providing frameworks for individual companies to train or code AI models themselves, towards providing  centrally trained models which address wider ranges of use cases \citep{maslej_artificial_2024}. The chance that an available model or service can address a given task out-of-the-box, as measured by leading benchmarks, is continually rising   \citep{reuel_betterbench_2024}. Even where there is customization, starting from such common frameworks necessarily generates interdependencies \citep{evans_2023_WeCollaborateWhatWeDesign,farrell2023underground}. Similarly these providers are now introducing versions of their  models allowed to be `agentic' --- to procure actions from other digital services directly without intermediation of a humman, or even to use voice or text interfaces to procure services from humans employed externally to the organization. The way such systems chain model calls and integrate access to external tools and systems is frequently described as at least partially autonomous \citep{durante_agent_2024, chan_2023_HarmsIncreasinglyAgenticAlgorithmicSystems}. This potentially breaks the call-and-response interaction pattern familiar from previous models, and with it, arguably Heidegger's idea of human application \citep[though see also][]{evans_2023_WeCollaborateWhatWeDesign}.

A further technical challenge is the micro-level intransparency of AI systems. Fully understanding AI systems on a mechanical level has long been infeasible due to the size, complexity, and intransparent learning mechanisms of the underlying models \citep{lipton_mythos_2017}. Recent growth in model size, capacity, and architectural complexity has exacerbated this; furthermore, business models providing models behind APIs, with the functioning hidden from the end user, are now commonplace \citep{burrell_how_2016, maslej_artificial_2024}. The value chain around AI systems is also very complex, often including human developers, data generators, procurers, deployers, users, and overseers across several organizations. This further complicates long-standing questions of responsibility attribution \citep{nissenbaum_accountability_1996,Santoni18,brown_allocating_2023, widder_dislocated_2023}. 

Such concerns --- and complexity --- are not unique to AI. Rather, they are also familiar critiques of governments and other large organizations, such as our earlier introduction to bureaucracy addresses. Thus claims that AI may potentially reduce, rather than enhance, bureaucratic rationality, or that the unclear inner workings of AI systems prevent the full attribution of responsibility for processes in which they are involved, need to be considered in context to a long history of similar complaints for the governments they supposedly obscure.


While concerns about transparency in both cases are valid, careful design of oversight and evaluation can mitigate them \citep{BrysonTheodorou19}. As \citet{sobel_legitimacy_2024} argues, the desiderata that explanations must fulfill to guarantee process legitimacy can be met without detailed technical understanding of the underlying systems, AI or otherwise. These desiderata include robustness to perturbation, causal attribution, and identification of human contributions to an outcome. Each of these is enabled by a disambiguation between mechanistic understanding and functional evaluation of AI systems \citep{doshi-velez_towards_2017,Santoni18}. 

Functional evaluation --- human verification that a given system is working correctly, on the terms of the organization or process within which it is embedded --- remains feasible. As an illustrative example, an AI system highlighting disputed points in two documents may be mechanically intransparent, but a human can verify via the model's inputs and outputs whether the points are correctly identified. \citet{doi:10.1177/09520767231197801} name further examples of translations into organizationally rational processes, such as the human generation of decision notification letters.

Indeed, oversight of agents and departments without exhaustive functional understanding is one variety of the bounded rationality --- the ability to act in limited time or on imperfect information --- that \citet{simon_administrative_1947} attributes to bureaucrats. Under the loose assumption that bureaucracies continue to take human-legible legislation as input and produce human-parseable outputs, functional evaluation is therefore always feasible in bureaucratic contexts. 

The same reasoning can be applied at smaller levels of granularity to design criteria and procurement demands for specific AI systems. This again allows bureaucrats to be meaningfully responsible. It is therefore not only a moral imperative \citep{Veluwenkamp03102024}, but also a technical possibility to create oversight structures which keep accountability for the outputs of AI systems solidly in the hands of humans, or at least of legal entities \citep{schmitz-etal-2025-oversight}. The EU AI Act is one example of regulation built on exactly this axiom \citep{popa_multilevel_2024}. The law of the European Union requires model oversight that is exhaustive enough to attribute responsibility to different human and institutional parties.

Such law is necessary. The required degree of capacity for oversight is not a given for AI systems, but rather a potential feature of any sociotechnical system. As such, it requires procurement, design, and verification, like any other feature. It is equally feasible that systems are designed to neglect or actively inhibit transparency --- preventing, rather than enabling, full or just responsibility distribution. \citet{elish_2019_MoralCrumpleZonesCautionaryTales} specifically cautions against humans overseeing AI being used as ``moral crumple zones" for the outcomes of AI systems which were in fact unsteerable by the supposed moral agent, or the actions of which cannot be functionally evaluated. Ensuring that oversight is technically feasible is therefore a technical design decision. But note such a design decision or law only {\em enables} just responsibility distribution, rather than constituting it in itself.

The interaction of human bureaucrats with AI systems cannot generally be construed simply as tool use. Yet despite the intransparency of full technical details of AI system operation, careful design of oversight enables full attribution of responsibility to humans. It does not restore the applicability of \textit{Zuhandenheit} \citep{heidegger_being_2008}, which would validate this attribution as the \textit{correct} relationship between humans and AI systems. The features of modern AI systems --- autonomy, generality, and deep integration into processes --- undermine this.  Nevertheless, such uncertainty is familiar to bureaucracies, given that details of even individual human behavior generation are also opaque. To solve the question of responsibility attribution in bureaucracies, two options remain: formulate another justification for keeping responsibility in human hands, or attribute responsibility to the AI systems themselves. We consider the latter option next.   

\begin{figure}[h]
    \centering
    \includegraphics[width=0.8\columnwidth]{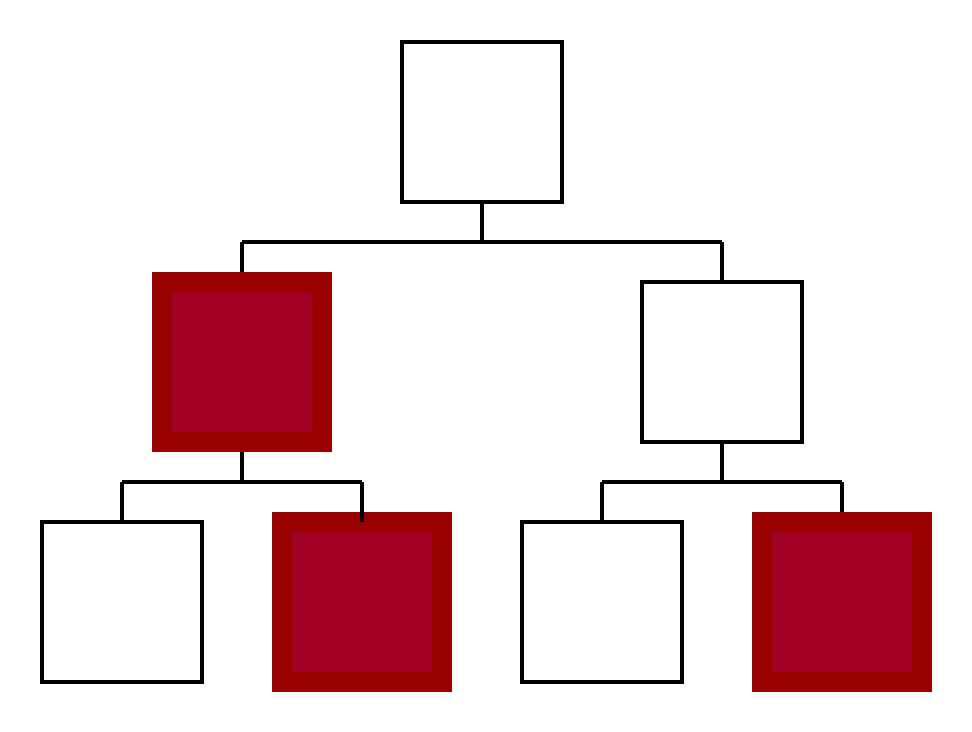}
    \captionof{figure}{Illustration of AI integration in bureaucracy under the assumption that AI systems can be moral subjects. Here, AI systems can be directly integrated in structures of responsibility attribution.}
    \label{fig:moralsubjects}
\end{figure}

\subsection{AI Systems as Moral Subjects} 

AI systems richly and obviously interact with the moral agency of individual bureaucrats \citep{bullock_2019_ArtificialIntelligenceDiscretionBureaucracy}. The possibility that they may be moral agents is therefore not only convenient for responsibility attribution, but arguably also intuitive. In this subsection, we argue that this possibility cannot be entertained because the required attribute of candidate moral agents, at least as defined here, 
is not guaranteed or even plausible in digital systems. We then address two alternative accounts of moral status --- the decoupling of moral agency and responsibility \citep{floridi_2004_MoralityArtificialAgents}, and the social attribution of moral agency \citep{coeckelbergh_2010_RobotRightsSocialRelationalJustificationMoral}. We argue that misapplication of either of these may cause ethics sinks (as defined earlier), or masked responsibility gaps \citep{matthias_responsibility_2004}. We arrive at the conclusion that bureaucracies must integrate AI systems into complete networks of human moral agency, despite AI marketing and features increasingly teasing the attribution of moral agency.

As derived in Section~\ref{dual.sec}, any conception of moral agency relevant for integration in bureaucracies is contingent on the candidate entity's susceptibility to dissuasion.  The aversions humans have to loss of status, loss of physical integrity, loss of liberty etc. are permanent, intrinsic, and systemic. These characteristics --- coupled with an ability to understand and negotiate legislative and other responsibilities --- allow entities to be integrated into structures of responsibility attribution that rely on reward and punishment, such as bureaucracies. The characteristics of reliable, long-term persuasion are shared with many other social animal species, the capacity to understand text may (arguably) be shared with some AI systems, but the conjunction of these two is unique to societies of adult humans who share sufficient culture to effectively communicate on such themes.

Much of the philosophical conversation on whether AI systems qualify for this definition of moral agency is perhaps over-focused on concepts like consciousness and pain --- for which there are no agreed, clear measures, or even medical definitions aligned with everyday intuitions \citep{dennett_1978_WhyYouCantmakecomputer,torrance_2014_ArtificialConsciousnessArtificialEthicsRealism,bishop_2009_WhyComputersCantFeelPain,sebo_2023_MoralConsiderationAIsystems2030}. 
Fortunately, such concerns are not particularly relevant to the desiderata just stated. AI systems are easily designed to respond to incentives, but just as easily designed to halt such responses. Digital systems can be upgraded, assaulted, or simply instructed to change their characteristics. Their memory relevant to aversion can be entirely and quickly replaced or deleted.  AI systems therefore do not afford the assumption that they are moral agents.

Even were a system to be developed where apparent self-motivation were intrinsic and inviolable, it is highly unlikely such a system would be a suitable candidate component for a government bureaucracy, for the reasons described in Section~\ref{dual.sec} concerning plurality and stewardship. Enforcement against humans --- and more importantly, collaboration and balance between humans making up a bureaucracy --- is possible because (and to the extent that) humans are roughly of the same power scale as each other, and have similar motivations rooted in our biological `hardware.' 

Digital systems by their nature can be scaled and altered in ways and timescales entirely unlike humans. Crucially, those timescales include the nearly-instantaneous. While present corporate practice has generated models that take months to retrain in full, the necessities of the market ensure that such models are firstly enormous and rare, and secondly very quickly redirectable though secondary learning systems. As such, digital systems are not candidate partners for processes underlying cooperation between humans \citep{evans_2023_WeCollaborateWhatWeDesign}. Further, to the extent they are sufficiently costly to be candidate for aversion, they are for the same reasons not candidate for plurality. The faux personality plurality created through secondary training is of no assistance here. Rather, it demonstrates again the highly transient and therefore unreliable nature of digital system performance.




We argued in the context of legitimacy and legislation that there is no way to separate moral agency from responsibility, nor legitimacy from the plurality of human opinions. In contrast, \citet{floridi_2004_MoralityArtificialAgents} argue that moral agency, particularly for digital moral agents, need not imply responsibility. \citeauthor{floridi_2004_MoralityArtificialAgents} propose a scope for moral agents with `mind-less morality' (p. 2), which can be \textit{accountable} for moral actions without being \textit{responsible} for them. 
They propose forms of censureship for misbehaving digital moral agents that are consistent with this definition (p. 20): monitoring and modification, removal to a disconnected component of cyberspace, and deletion from cyberspace 
\citep[cf.][]{popa_multilevel_2024}.  It should be evident that such penalties are not intended to constitute means of dissuasion or punishment towards which the AI agent itself feels aversion. Rather, as \citeauthor{floridi_2004_MoralityArtificialAgents} state, these are simply take-downs, identical to the workings of cybersecurity systems. Although unquestionably a matter of definition, we see little value in referring to fly swatting as attribution of responsibility, or to flies as moral agents.

If looking at such penalties from an anthropomorphic or even zoomorphic lens, it might seem that the fear of deletion should necessarily motivate an agent. This forgets that agents need have no concept available along these lines---no index to update to increase `fear' or vigilance, no identity to share with missing systems. These are program options, meaning they can be inserted or deleted at will not only by system developers, but by system operators, and by cyber-capable opponents. The structural implications of \citet{floridi_2004_MoralityArtificialAgents} model therefore do not contradict ours. In both, humans, who both \textit{can} and \textit{must} take responsibility, are equipped with means to steer and oversee the artificial agents which cannot \citep{floridi_ai_2025}.

A second possible objection is attribution of moral agency not based on any objective measures of an entity, but based on how it is perceived socially. In this ``social relationist'' conception, society’s duty of care to a candidate moral subject is defined by social attribution, treatment, and consensus \citep{coeckelbergh_2012_GrowingMoralRelationsCritiqueMoral}. Entities are granted patiency when society perceives them to be worthy of it. Moreover, if individuals identify with such entities, perceived harm to the entity could be realized as harm to the self. 

It would follow from this conception that whether or not AI systems are attributed patiency is a direct consequence of their design, particularly of the interfaces humans have to AI systems. These interfaces are the primary source we can draw on when making inferences of moral status \citep{gray_2012_MoralityTakesTwoDyadicmorality}. While such a heuristic for attributing patiency may have in previous centuries been useful for recognizing both human universals and animal welfare \citep{KantCollins85}, in the context of modern corporate capacities for attention and sympathy generation, it becomes evidently weaponizable. 

Indeed, there is increasing evidence that humans do ascribe moral statuses to AI systems based on interactions \citep{ladak_2025_RobotsChatbotsSelfDrivingCarsPerceptions}, including in public-sector contexts \citep{shank_2018_AttributionsMoralityMindartificialintelligence}. Particularly anthropomorphic design, such as human-like emotion expression or physical features, has outsize influence \citep{ladak_2024_WhichArtificialIntelligencesPeopleCare, yam_2022_WhenYourBossrobotWorkers}.  It is therefore plausible that social relationist conceptions of moral patiency most closely describe how this status is presently attributed to AI systems in practice. 

A moral issue results: a society might ascribe patiency to commercial products based on the effectiveness of a corporate advertising budget, rather than on real impacts either on the systems' intended purpose, or even on the system itself. In the case we are considering here, of systems with agency in governments, this is clearly unacceptable. For this reason, AI `soft law' starting with the UK's Principle's of Robotics  \citep{ConSciPoR11}, continuing with the OECD's Principles of AI \citep{OECDAI19}, culminating in the UNESCO Recommendation on the Ethics of Artificial Intelligence \citep{UNESCOairec22}, and then realized into law by e.g. the European Union's AI Act \citep{eu2024aiact} all prohibit deceptive engineering that encourages falsely identifying AI as human-like.


Arguments such as the above are the sorts of things that can cause the ethics sinks we set out to avoid in the introduction --- structures that allow the dissipation of responsibility. More precisely, an ethics sink might be thought of as a responsibility gap \citep{matthias_responsibility_2004}  \textit{masked} by the mistaken attribution of moral agency to an inappropriate entity. Two evident potential causes for such mistaken attribution then could be misunderstanding of simple measures of censure as measures of punishment with dissuasive efficacy \citep{floridi_2004_MoralityArtificialAgents}, or a misattribution of moral efficacy (a potential for moral agency) based on social inference \citep{gray_2012_MoralityTakesTwoDyadicmorality}. 

Philosophers have long used fictive imaginings concerning AI as `intuition pumps' for examining aspects of the human condition. But now that we often discuss real systems that (among other things) alter the performance and security of public services, and therefore impact human rights and welfare, it is time to clearly separate fiction from facts. Design decisions, whether of AI systems themselves or of moral ontologies for describing them, that may impact the probability of misattributed moral agency are themselves moral acts. While we do not want to limit academic inquiry or corporate innovation, we also want our government to work, and our electorates to be able to appropriately assess and monitor whether their governments are working for them. 

Ethics sinks evidently undermine the legitimacy of bureaucratic institutions, as they disrupt full and just attribution of responsibility. A salient variation of this is the expectation of explanations for decisions, which is commonly cited as a prerequisite for integration of AI systems into bureaucratic decision-making processes \citep{abu_elyounes_computer_2020, de_boer_automation_2023, bell_replacing_2021}. As \citet{sparrow_2025_TestimonyGapMachinesReasons} argue, even the giving of reasons for a decision can be construed as a moral act. As process legitimacy indeed requires the organization as a whole to produce explanations \citep{sobel_legitimacy_2024}, deference of explanatory power to systems which are misattributed moral agency is therefore a credible risk.

In summary, if we use the term `moral agent' as the attributionists would prefer, to indicate those entities that should be held to account --- that should be designated as responsible, and then we assume as most functionalist moral philosophers do, that the purpose of accountability and responsibility is maintenance or improvement of the society, then only humans should be held as moral agents. Neither digital systems, nor non-human animals are adequate candidates. Digital systems are both too impermanent and too entangled with large scale corporations and other such infrastructure to be subject to persuasion of dissuasion. Animals with whom we cannot converse we can also not negotiate with and form the sort of peer networks required to run significantly organizations in a stable manner, even if we can to a limited extent integrate them as members of e.g. households \citep{bryson_2018_PatiencyNotVirtuedesignintelligent}.

But digital systems in themselves can never be rationally treated as either legal or moral agents. Systems of responsibility allocation designed around AI systems --- including bureaucracies --- therefore cannot assume the moral agency of these systems, or present them as human-like trustworthy entities. Consequently, the primary purpose of these mechanisms must be to allocate responsibility for processes and decisions involving AI systems fully to legitimate moral agents.

\begin{figure}[h]
    \centering
    \includegraphics[width=0.8\columnwidth]{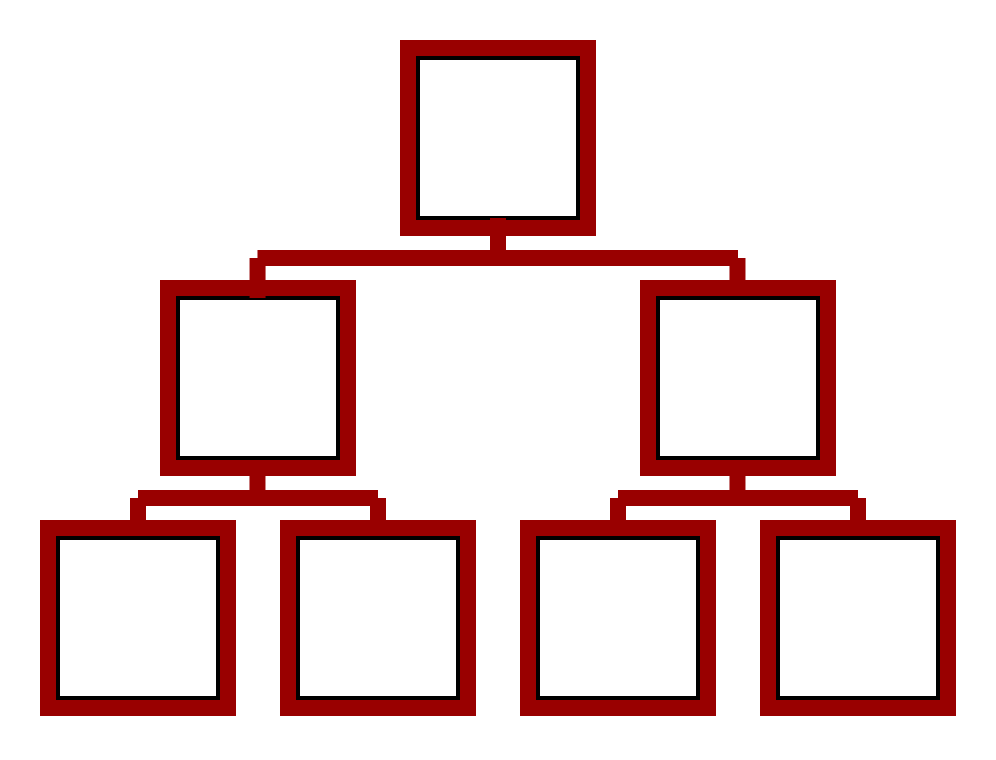}
    \captionof{figure}{Our conception of AI systems as ``muscles, sinews and ligaments" of a bureaucracy. AI systems increase the agency of the organization as a whole by amplifying the capacities of human moral agents --- but do not negate the distribution of responsibility exclusively among the legitimate moral agents. In other words, AI integrates into the existing role of the institutional structure of a Weberian bureaucracy.}
    \label{fig:infrastructure}
\end{figure}

\subsection{AI as Augmenting Infrastructure} 

The previous two subsections appear to result in a contradiction: on the one hand, because AI systems are not mere tools, the ``tool conception'' cannot be used to justify attributing all responsibility for their effects to humans. On the other, attributing moral agency to any digital system results in ethics sinks --- AI systems themselves should not be attributed responsibility because they cannot be held to account. Where, then, does responsibility go? Here, we argue that while responsibility must indeed ultimately lie with humans, AI systems do alter how that responsibility is understood, attributed, and distributed. There is ample room for a third role: non-moral artifacts which nonetheless interact with the moral agency of humans. We argue this exact status is already held by the institutional structure of bureaucracy, and conceptualize AI systems as an extension thereof. AI systems can be the muscles, sinews, and ligaments --- smart organs connecting the orientation of an institution with its actions. 

It should now be apparent that AI systems impact a bureaucracy's ethics solely by altering the capacities of the humans that root its core moral agency \citep{bullock_2019_ArtificialIntelligenceDiscretionBureaucracy}. This should come as no surprise: digital instruments have a well-established capacity to alter the agency of human civil servants. Arguably the core insight of the models of system- and screen-level bureaucracy \citep{bovens_street-level_2002}, is that spaces of discretion are ceded by individual civil servants to IT systems. Presently, where these systems are introduced, discretionary space tends to be centralized down to a relatively small number of system designers and data analysts, rather than distributed among large numbers of street-level bureaucrats. This process, without adequate checks and care, can lead to a de-individualization not just of citizens interacting with PSOs, but of the human civil servants within them \citep{zouridis_automated_2019}. 

Beyond restricting discretion, AI systems are increasingly used to influence \textit{how} discretion is exercised within any remaining discretionary spaces \citep{bell_replacing_2021, bullock_2020_ArtificialIntelligenceBureaucraticformdiscretion}. In a phenomena known as `automation bias,' many users including some bureaucrats tend to trust AI outputs indiscriminately. This can lead to qualitative differences in evaluations by increasing throughput, and introduce myriad and arbitrarily-expensive forms of bias, prejudice, or simple falsehoods \citep{eubanks_automating_2018,Wallis21,Perry23, alon-barkat_2023_HumanAIInteractionsPublicSector}. AI systems that are anthropomorphic --- communicating in natural language, for example --- are more likely to influence human discretion, almost as a form of conformity bias \citep{Troshani03092021}. Referring to working with such systems as `collaboration' further confuses and indeed compromises the bureaucrat's true agency \citep{evans_2023_WeCollaborateWhatWeDesign}.


There are no simple solutions to such quandaries, yet fortunately a variety of civilizations have spent millennia developing complexes of administrative solutions which academic disciplines such as Public Administration work to research and advance. Weberian bureaucracy is intended to ensure formal rationality through a variety of means that both constrain yet augment human agency, power, and ability. Means include formal procedure, legal norms, chains of hierarchy and delegation, documentation and record-keeping \citep{weber1978economy, clegg_2009_WeberSinteringIronCageTranslation}. Again, such means evidently hold no moral agency of their own, though arguably as useful artefacts they may hold a form of patiency in service of the humans who could benefit from that use \citep{KantCollins85}. Infrastructure is not a receptacle of responsibility, rather it augments that of the human constituents of the bureaucracy. 

In particular, as detailed above, institutions facilitate not only diffusion but also reattribution of responsibility. This is what enables the public service bargain, and the useful fiction of attribution of responsibility to known, public, elected figures. AI systems too should be specified, procured, and integrated to facilitate this legitimacy. It is not necessary for AI systems to differ substantially from any other structural means for framing and understanding responsibility with respect to human bureaucrats and legislators. 
Much like procedure, rules, and precedent may be deferred to by a civil servant in justifying their actions, so may the outputs of an AI system. In neither case should this imply any abolition of responsibility, though in either case design choices are also available to entirely obscure responsibility. The choice to create legibility \textit{is} a choice, and a core one to effective bureaucracy, and a stable, legitimate state.

We arrive then at a conception of AI that --- if well designed and integrated --- can be part of the ``muscles, sinews, and ligaments" of the bureaucratic organization. Procured AI systems should be designated to be elements that increases the agency and structural capacity of the system as a whole, without taking any role in its orientation. Orientation is due to the discretion of individual bureaucrats, in a context ultimately grounded in the decisions of voters. This conception is already afforded by analogy to, or embedding in, extant institutional structures of bureaucracy. 

Beyond resolving the false dichotomy between tools and moral agents, our conception of AI as Weberian bureaucratic infrastructure should also prove more future-proof than either of the earlier  models. Even intuitively, AI that is increasingly general, ubiquitously embedded in systems and processes, and agentic may be more convincingly conceptualized as a part of the institutional framework. 

\vspace{3mm}

\begin{tcolorbox}[
  colback=blue!10!white,
  colframe=blue!70!black,
  title=Moral Agency Framework,
  rounded corners,
  boxrule=0.8pt,
  fonttitle=\bfseries\color{white},
  colbacktitle=blue!70!black,
  title style={fill=blue!80!black, left=2mm, right=2mm, top=3mm, bottom=3mm},
  before upper={\setlength{\leftskip}{0pt}},
  before upper*={\setlist[itemize,1]{left=0pt}\setlist[enumerate,1]{left=0pt}},
]
\begin{enumerate}
    \item \textbf{Maintain clear and just human lines of accountability} for any decisions involving or affected by AI systems.
    \item \textbf{Ensure that humans can verify} the correctness and appropriateness of AI system outputs within their operational context.
    \item \textbf{Introduce AI only where it does not inhibit} a bureaucracy's ability to pursue legitimacy and stewardship.
\end{enumerate}
\end{tcolorbox}

\section{A Moral Agency Framework} \label{fw.sec}

How can we ensure the continuation of legitimacy and stewardship aspects of governance as we procure and integrate AI into its processes? This section proposes one possible moral agency framework for achieving those goals. Our framework  rests on our conception of a bureaucracy as a linked set of human moral agents. We refer to such agents as ``legitimate'' to the extent that their accountability can be meaningfully traced back through the web of responsibility to those answerable to the public the government serves. AI systems, vendors, consultancies, and any other agencies should all fall clearly into the responsibility basin of some such legitimate agent. As just explained, AI takes a role comparable to the muscles and ligaments of a human body: increasing the agency of the system overall, while not determining its goals nor deciding between contested alternative policies. Earlier sections sought to clarify that this state is achievable, subject to design decisions. With this framework, we seek to guide design decision towards that goal.\\

\subsection*{1. Maintain clear, just human lines of accountability.}

We established that (a) bureaucracies must attribute all responsibility to moral agents to retain their democratic legitimacy, and (b) the moral agency of AI systems themselves can never be assumed, despite their impact on the moral agency of humans. Therefore, in order to avoid ethics sinks, bureaucracies must maintain structures of accountability that attribute all responsibility for organizational outcomes to human moral agents, who must have both discretionary space and aversion to punishment \citep[as opposed to those constituting moral crumple zones, who have only the latter, cf.][]{elish_2019_MoralCrumpleZonesCautionaryTales}. This includes responsibility for actions taken with the aid of AI systems, and any other contributions towards organizational outputs which introduce ambiguity in responsibility. Examples besides the output of IT systems include logistic supply chains, systems of vendors, and outputs of consultancies.

This is not a novel contribution in itself, but a restatement of the merits of formal rationality as per \citet{weber1978economy}. Indeed, many of Weber's signature attributes of bureaucracy fulfill exactly this purpose. Our novel contribution is the assertion that this remains feasible, and normatively desirable, even when AI systems are introduced.

\subsection*{2. Ensure humans can verify correct functioning of the systems they deploy.}

Attribution of responsibility to legitimate actors is a core purpose of bureaucracy, and the core purpose of this framework. Transparency is definitionally what enables this attribution. Transparency allows legitimate agents the level of insight required to understand whether a given system is performing correctly, and therefore allows the legitimate agent to take on responsibility for the outputs of the systems intended to augment their agency. It is not possible to legitimately attribute credit or blame to humans for outputs a human cannot themselves verify as correct or wrong, at least within some prespecified range of certainty. This does not imply, of course, that a given human's understanding of an AI system's function immediately implies that human's responsibility for the system's actions. It merely qualifies the outputs of that system for consideration in structures of responsibility attribution. In so doing, it enables the functioning of these structures --- for example by allowing the more granular attribution of responsibility across complex AI value chains.

We establish in Section~\ref{sit.sec} that functional evaluation of AI systems, to the level required for this responsibility attribution, is readily technically feasible today, and will remain feasible as long as we require our bureaucracies to have human-legible inputs (legislation) and outputs (policy outcomes) on an institutional level. It is therefore a design decision --- on both technical and institutional levels --- whether bureaucratic structures afford this functional legibility. In the European Union, this requirement is legally mandated for all `high risk' AI systems --- that is those that impact human lives significantly --- by the AI Act \citep{eu2024aiact}.

\subsection*{3. Introduce AI only where it does not inhibit either of bureaucracy's dual aims: legitimacy and stewardship.}

Modifications to the agency of legitimate actors are a key mechanism by which bureaucracies alter the balance between their twin aims: legitimacy, the faithful implementation of legislative will, and stewardship, the provision of long-term stability. Integration of AI systems may disturb this balance. For example, AI may increase the efficiency of a process, but centralize and standardize decision-making in a way that decreases the organization's resilience with respect to legislative impulses, thereby threatening its ability to provide stewardship  \citep{newman_2022_DigitalTechnologiesArtificialintelligencebureaucratic}. It is unlikely without design that a given alteration improves a bureaucracy's capacity to fulfill both these aims simultaneously, given some inherent trade-offs between them. 

However, it is reasonable to require that any introduction of AI that moves an organization towards one of these goals does not jeopardize the organization's ability to fulfill the other. This is readily possible in practice: public organizations are generally far from the Pareto frontier of agency and stability (among other public values) \citep{grandy_efficient_2009}. That is, they are generally suboptimal in both these dimensions, owing to well-documented administrative inefficiencies such as growth beyond optimal size, budget maximization, and issues of information capacity \citep{simon_administrative_1947, lindblom_science_1959, niskanen_bureaucracy_1971}. The conceptual trade-off we describe above must therefore rarely be invoked in practice, because room remains for improvements in one dimension that do not impede the other. It might also be that with careful design and good intention, some improvements might facilitate the same legitimate agents being facilitated in both goals.

\section{Conclusion: It’s the Design Decisions}

Careless integration of AI into bureaucracies may cause threats to the mechanisms that make these bureaucracies legitimate or effective institutions. None of our suggestions here are a given. Anyone can easily design a digital system that expunges all record-keeping, centralizes all control, or limits internal communication or innovation. These threats, coupled with confusion about the moral status of AI systems, have led to calls to avoid adopting AI systems in bureaucracies outright, or only with extreme caution \citep{cetina2024adoption}. 

We hereby disagree. We hope we have established in this work that it is readily possible to legitimately integrate AI into government bureaucracies, despite the fact that AI is neither exactly a simple tool nor itself a moral agent. On both technical and institutional levels, the means exist to ensure structures of responsibility attribution --- which make bureaucracies legitimate enactors of legislative will --- remain intact when AI systems are introduced. Bureaucratic design can require that all responsibility for outputs of the institution continues to rest with legitimate agents. Further, it is even possible for well-implemented AI systems to \textit{increase} the formal rationality, transparency, and legitimacy of bureaucracies, for example by increasing clarity and comprehensiveness of data and processes, and by automating record keeping. Such design choices should not be hindered by naive debates about moral consideration of AI systems, nor of AI's alleged inherent intransparency. AI systems can be made transparent at the levels necessary for human control. It is possible to legitimately realize the enormous potential of AI to increase the capacity of bureaucracies towards their aims.

Having said that, we emphasize again that such good outcomes are by no means a given. Many well-known vendors are evidently attempting to increase anthropomorphic trust in their products, despite empirical studies showing such sensations of trust lead to inferior, even dangerous (mis)use of AI \citep{Perry23}. Our optimism for the capacity for the legitimate and transparent application of AI to government bureaucracies must therefore \textit{not} be mistaken for an unconditional support for all endeavors at digitalization. Misapplication of AI may lead to the weakening of public-sector organizations' ability to perform self improvement, which has been suggested as a possible contributing factor to the bureaucratic support and implementation of atrocities, also known as ``administrative evil'' \citep{young_2021_ArtificialIntelligenceAdministrativeEvil}. Maintaining responsibility and internal capacity for such self improvement is of a piece with wider stewardship. We hope the means such as we have advised here may be sufficient to avoid such outcomes \citep{senge_fifth_1990, argyris_organizational_1997}. 

Further, the `iron cages' of a strong bureaucracy may necessarily facilitate rapid course correction for a state, allowing the expression of stewardship of new governments with `evil,' even self-harming aims \citep{Heldring23}. If so, the inherent surveillance and transparency capacities of AI may always be at risk of misdeployments aimed solely at strengthening the iron of the bureaucrats' cages, enforcing conformity to the government's mandate. An essential part of ensuring our framework guarantees good outcomes may require careful monitoring or even red-teaming to ensure that a new government or invading force with ill will towards the population could not rapidly transform the system into one in opposition of legitimate and constitutional aims.

The challenge, then, is not to resist AI integration in principle, but to take seriously the design responsibilities that come with it. If done well, such integration can not only preserve but actively deepen both the capacity and the legitimacy of public institutions. The legitimacy of AI integration in bureaucracies rests entirely on human design decisions: design procurement, deployment, training and integrating AI systems; design of oversight structures and interaction processes for functional evaluation; design of the processes that AI systems are embedded in; design of the responsibility attribution structure around the AI systems.

\clearpage
\onecolumn
\bibliographystyle{apalike-nonote}
\bibliography{master,name}

\begin{thebibliography}{}

\bibitem[Abu~Elyounes, 2020]{abu_elyounes_computer_2020}
Abu~Elyounes, D. (2020).
\newblock '{Computer} {Says} {No}!': {The} {Impact} of {Automation} on the {Discretionary} {Power} of {Public} {Officers}.

\bibitem[ACSH, 2023]{acsh_stewardship_2023}
ACSH (2023).
\newblock Stewardship and {Public} {Service}: {An} {Introduction}.
\newblock Technical report, Astana Civil Service Hub.

\bibitem[Alon-Barkat and Busuioc, 2023]{alon-barkat_2023_HumanAIInteractionsPublicSector}
Alon-Barkat, S. and Busuioc, M. (2023).
\newblock Human–{AI} {Interactions} in {Public} {Sector} {Decision} {Making}: “{Automation} {Bias}” and “{Selective} {Adherence}” to {Algorithmic} {Advice}.
\newblock {\em Journal of Public Administration Research and Theory}, 33(1):153--169.

\bibitem[Argyris and Schön, 1997]{argyris_organizational_1997}
Argyris, C. and Schön, D.~A. (1997).
\newblock Organizational {Learning}: {A} {Theory} of {Action} {Perspective}.
\newblock {\em Reis}, (77/78):345--348.

\bibitem[Barez et~al., 2025]{barez_toward_2025}
Barez, F., Friend, I., Reid, K., and Krawczuk, I. (2025).
\newblock Toward {Resisting} {AI}-{Enabled} {Authoritarianism}.

\bibitem[Bell, 2021]{bell_replacing_2021}
Bell, B.~W. (2021).
\newblock Replacing {Bureaucrats} with {Automated} {Sorcerers}?
\newblock {\em Daedalus}, 150(3):89--103.

\bibitem[Bentham, 2005]{bentham_2005_IntroductionPrinciplesMoralsLegislation}
Bentham, J. (2005).
\newblock {\em An {Introduction} to the {Principles} of {Morals} and {Legislation}}.
\newblock Oxford University Press, Oxford.

\bibitem[Bishop, 2009]{bishop_2009_WhyComputersCantFeelPain}
Bishop, M. (2009).
\newblock Why {Computers} {Can}’t {Feel} {Pain}.
\newblock {\em Minds and Machines}, 19(4):507--516.

\bibitem[Boden et~al., 2011]{ConSciPoR11}
Boden, M., Bryson, J., Caldwell, D., Dautenhahn, K., Edwards, L., Kember, S., Newman, P., Parry, V., Pegman, G., Rodden, T., Sorrell, T., Wallis, M., Whitby, B., and Winfield, A. (2011).
\newblock Principles of robotics: regulating robots in the real world.
\newblock {\em Connection Science}, 29(2):124--129.

\bibitem[Bovens and Zouridis, 2002]{bovens_street-level_2002}
Bovens, M. and Zouridis, S. (2002).
\newblock From {Street}-{Level} to {System}-{Level} {Bureaucracies}: {How} {Information} and {Communication} {Technology} is {Transforming} {Administrative} {Discretion} and {Constitutional} {Control}.
\newblock {\em Public Administration Review}, 62(2):174--184.

\bibitem[Brown, 2023]{brown_allocating_2023}
Brown, I. (2023).
\newblock Allocating accountability in {AI} supply chains.
\newblock Technical report, Ada Lovelace Institute.

\bibitem[Bryson, 2018]{bryson_2018_PatiencyNotVirtuedesignintelligent}
Bryson, J.~J. (2018).
\newblock Patiency is not a virtue: the design of intelligent systems and systems of ethics.
\newblock {\em Ethics and Information Technology}, 20(1):15--26.

\bibitem[Bryson et~al., 2017]{bryson_2017_PeopleLegalLacunasyntheticpersons}
Bryson, J.~J., Diamantis, M.~E., and Grant, T.~D. (2017).
\newblock Of, for, and by the people: the legal lacuna of synthetic persons.
\newblock {\em Artificial Intelligence and Law}, 25(3):273--291.

\bibitem[Bryson and Theodorou, 2019]{BrysonTheodorou19}
Bryson, J.~J. and Theodorou, A. (2019).
\newblock How society can maintain human-centric artificial intelligence.
\newblock In Toivonen-Noro, M. and Saari, E., editors, {\em Human-Centered Digitalization and Services}, pages 305--323. Springer, Berlin.

\bibitem[Bullock et~al., 2020]{bullock_2020_ArtificialIntelligenceBureaucraticformdiscretion}
Bullock, J., Young, M.~M., and Wang, Y.-F. (2020).
\newblock Artificial intelligence, bureaucratic form, and discretion in public service.
\newblock {\em Information Polity}, 25(4):491--506.

\bibitem[Bullock, 2019]{bullock_2019_ArtificialIntelligenceDiscretionBureaucracy}
Bullock, J.~B. (2019).
\newblock Artificial {Intelligence}, {Discretion}, and {Bureaucracy}.
\newblock {\em The American Review of Public Administration}, 49(7):751--761.

\bibitem[Bullock et~al., 2025]{bullock_2025_AGIGovernmentsFreeSocieties}
Bullock, J.~B., Hammond, S., and Krier, S. (2025).
\newblock {AGI}, {Governments}, and {Free} {Societies}.

\bibitem[Burrell, 2016]{burrell_how_2016}
Burrell, J. (2016).
\newblock How the machine ‘thinks’: {Understanding} opacity in machine learning algorithms.
\newblock {\em Big Data \& Society}, 3(1):2053951715622512.

\bibitem[Cetina~Presuel and Martinez~Sierra, 2024]{cetina2024adoption}
Cetina~Presuel, R. and Martinez~Sierra, J.~M. (2024).
\newblock The adoption of artificial intelligence in bureaucratic decision-making: A weberian perspective.
\newblock {\em Digital Government: Research and Practice}, 5(1):1--20.

\bibitem[Chan et~al., 2023]{chan_2023_HarmsIncreasinglyAgenticAlgorithmicSystems}
Chan, A., Salganik, R., Markelius, A., Pang, C., Rajkumar, N., Krasheninnikov, D., Langosco, L., He, Z., Duan, Y., Carroll, M., Lin, M., Mayhew, A., Collins, K., Molamohammadi, M., Burden, J., Zhao, W., Rismani, S., Voudouris, K., Bhatt, U., Weller, A., Krueger, D., and Maharaj, T. (2023).
\newblock Harms from {Increasingly} {Agentic} {Algorithmic} {Systems}.

\bibitem[Clegg and Lounsbury, 2009]{clegg_2009_WeberSinteringIronCageTranslation}
Clegg, S. and Lounsbury, M. (2009).
\newblock Weber: {Sintering} the {Iron} {Cage} {Translation}, {Domination}, and {Rationality} {Stewart} {Clegg}.
\newblock {\em The Oxford Handbook of Sociology and Organization Studies: Classical Foundations}.

\bibitem[Coeckelbergh, 2010]{coeckelbergh_2010_RobotRightsSocialRelationalJustificationMoral}
Coeckelbergh, M. (2010).
\newblock Robot {Rights}? {Towards} a {Social}-{Relational} {Justification} of {Moral} {Consideration}.
\newblock {\em Ethics and information technology}, 12(3):209--221.

\bibitem[Coeckelbergh, 2012]{coeckelbergh_2012_GrowingMoralRelationsCritiqueMoral}
Coeckelbergh, M. (2012).
\newblock {\em Growing {Moral} {Relations}: {Critique} of {Moral} {Status} {Ascription}}.
\newblock Springer.

\bibitem[de~Boer et~al., 2023]{de_boer_automation_2023}
de~Boer, N., , and Raaphorst, N. (2023).
\newblock Automation and discretion: explaining the effect of automation on how street-level bureaucrats enforce.
\newblock {\em Public Management Review}, 25(1):42--62.

\bibitem[Dennett, 1978]{dennett_1978_WhyYouCantmakecomputer}
Dennett, D.~C. (1978).
\newblock Why you can't make a computer that feels pain.
\newblock {\em Synthese}, 38(3):415--456.

\bibitem[Doshi-Velez and Kim, 2017]{doshi-velez_towards_2017}
Doshi-Velez, F. and Kim, B. (2017).
\newblock Towards {A} {Rigorous} {Science} of {Interpretable} {Machine} {Learning}.

\bibitem[Durante et~al., 2024]{durante_agent_2024}
Durante, Z., Huang, Q., Wake, N., Gong, R., Park, J.~S., Sarkar, B., Taori, R., Noda, Y., Terzopoulos, D., Choi, Y., Ikeuchi, K., Vo, H., Fei-Fei, L., and Gao, J. (2024).
\newblock Agent {AI}: {Surveying} the {Horizons} of {Multimodal} {Interaction}.

\bibitem[Elish, 2019]{elish_2019_MoralCrumpleZonesCautionaryTales}
Elish, M.~C. (2019).
\newblock Moral {Crumple} {Zones}: {Cautionary} {Tales} in {Human}-{Robot} {Interaction}.
\newblock {\em Engaging Science, Technology, and Society}, 5:40--60.

\bibitem[Eubanks, 2018]{eubanks_automating_2018}
Eubanks, V. (2018).
\newblock {\em Automating {Inequality}: {How} {High}-{Tech} {Tools} {Profile}, {Police}, and {Punish} the {Poor}}.
\newblock St. Martin's Publishing Group.

\bibitem[{European Parliament and Council of the European Union}, 2024]{eu2024aiact}
{European Parliament and Council of the European Union} (2024).
\newblock {R}egulation ({EU}) 2024/1689 of the {E}uropean {P}arliament and of the {C}ouncil of 13 {J}une 2024 laying down harmonised rules on artificial intelligence and amending {R}egulations ({EC}) {N}o 300/2008, ({EU}) {N}o 167/2013, ({EU}) {N}o 168/2013, ({EU}) 2018/858, ({EU}) 2018/1139 and ({EU}) 2019/2144 and {D}irectives 2014/90/{EU}, ({EU}) 2016/797 and ({EU}) 2020/1828.
\newblock {\em Official Journal of the European Union, L Series}, (2024/1689).

\bibitem[Evans et~al., 2023]{evans_2023_WeCollaborateWhatWeDesign}
Evans, K.~D., Robbins, S.~A., and Bryson, J.~J. (2023).
\newblock Do {We} {Collaborate} {With} {What} {We} {Design}?
\newblock {\em Topics in Cognitive Science}, page tops.12682.

\bibitem[Farrell and Newman, 2023]{farrell2023underground}
Farrell, H. and Newman, A. (2023).
\newblock {\em Underground empire: How {A}merica weaponized the world economy}.
\newblock Random House.

\bibitem[Floridi, 2025]{floridi_ai_2025}
Floridi, L. (2025).
\newblock {AI} as {Agency} without {Intelligence}: {On} {Artificial} {Intelligence} as a {New} {Form} of {Artificial} {Agency} and the {Multiple} {Realisability} of {Agency} {Thesis}.
\newblock {\em Philosophy \& Technology}, 38(1):30.

\bibitem[Floridi and Sanders, 2004]{floridi_2004_MoralityArtificialAgents}
Floridi, L. and Sanders, J. (2004).
\newblock On the {Morality} of {Artificial} {Agents}.
\newblock {\em Minds and Machines}, 14(3):349--379.

\bibitem[Gailmard and Patty, 2012]{gailmard_2012_FormalModelsBureaucracy}
Gailmard, S. and Patty, J.~W. (2012).
\newblock Formal {Models} of {Bureaucracy}.
\newblock {\em Annual Review of Political Science}, 15(1):353--377.

\bibitem[Gasser and Mayer-Schönberger, 2024]{gasser_guardrails_2024}
Gasser, U. and Mayer-Schönberger, V. (2024).
\newblock {\em Guardrails: {Guiding} {Human} {Decisions} in the {Age} of {AI}}.
\newblock Princeton University Press.

\bibitem[Gay, 2000]{gay_praise_2000}
Gay, P.~d. (2000).
\newblock {\em In {Praise} of {Bureaucracy}: {Weber} - {Organization} - {Ethics}}.
\newblock SAGE Publications.

\bibitem[Grandy, 2009]{grandy_efficient_2009}
Grandy, C. (2009).
\newblock The "{Efficient}" {Public} {Administrator}: {Pareto} and a {Well}-{Rounded} {Approach} to {Public} {Administration}.
\newblock {\em Public Administration Review}, 69(6):1115--1123.

\bibitem[Gray and Wegner, 2012]{gray_2012_MoralityTakesTwoDyadicmorality}
Gray, K. and Wegner, D.~M. (2012).
\newblock Morality takes two: {Dyadic} morality and mind perception.
\newblock In {\em The social psychology of morality: {Exploring} the causes of good and evil}, Herzliya series on personality and social psychology, pages 109--127. American Psychological Association, Washington, DC, US.

\bibitem[Gunkel, 2018]{gunkel_other_2018}
Gunkel, D.~J. (2018).
\newblock The {Other} {Question}: {Can} and {Should} {Robots} {Have} {Rights}?
\newblock {\em Ethics and Information Technology}, 20(2):87--99.

\bibitem[Habermas, 2015]{habermas_between_2015}
Habermas, J. (2015).
\newblock {\em Between {Facts} and {Norms}: {Contributions} to a {Discourse} {Theory} of {Law} and {Democracy}}.
\newblock John Wiley \& Sons.

\bibitem[Heath, 2020]{heath_2020_MachineryGovernmentPublicAdministrationLiberal}
Heath, J. (2020).
\newblock {\em The {Machinery} of {Government}: {Public} {Administration} and the {Liberal} {State}}.
\newblock Oxford University Press.

\bibitem[Heidegger, 2008]{heidegger_being_2008}
Heidegger, M. (2008).
\newblock {\em Being and {Time}}.
\newblock HarperCollins.

\bibitem[Heldring, 2023]{Heldring23}
Heldring, L. (2023).
\newblock Bureaucracy as a tool for politicians: Evidence from germany.

\bibitem[Hinton et~al., 1987]{hinton1987learning}
Hinton, G.~E., Nowlan, S.~J., and {others} (1987).
\newblock How learning can guide evolution.
\newblock {\em Complex systems}, 1(3):495--502.

\bibitem[Hjaltalin and Sigurdarson, 2024]{hjaltalin_strategic_2024}
Hjaltalin, I.~T. and Sigurdarson, H.~T. (2024).
\newblock The strategic use of {AI} in the public sector: {A} public values analysis of national {AI} strategies.
\newblock {\em Government Information Quarterly}, 41(1):101914.

\bibitem[Hood and Lodge, 2006]{hood_2006_PoliticsPublicServiceBargainsReward}
Hood, C. and Lodge, M. (2006).
\newblock {\em The {Politics} of {Public} {Service} {Bargains}: {Reward}, {Competency}, {Loyalty} - {And} {Blame}}.
\newblock OXFORD UNIV PR, Oxford, illustrated edition edition.

\bibitem[Kant and Collins, 1785]{KantCollins85}
Kant, I. and Collins, G.~L. (1785).
\newblock {\em Moralphilosophie {C}ollins}, volume XXVII, pages 237--473.
\newblock Walter de Gruyter, Berlin, 1974 edition.

\bibitem[Ladak et~al., 2024]{ladak_2024_WhichArtificialIntelligencesPeopleCare}
Ladak, A., Harris, J., and Anthis, J.~R. (2024).
\newblock Which {Artificial} {Intelligences} {Do} {People} {Care} {About} {Most}? {A} {Conjoint} {Experiment} on {Moral} {Consideration}.

\bibitem[Ladak et~al., 2025]{ladak_2025_RobotsChatbotsSelfDrivingCarsPerceptions}
Ladak, A., Wilks, M., Loughnan, S., and Anthis, J.~R. (2025).
\newblock Robots, {Chatbots}, {Self}-{Driving} {Cars}: {Perceptions} of {Mind} and {Morality} {Across} {Artificial} {Intelligences}.

\bibitem[Latour, 2005]{latour_reassembling_2005}
Latour, B. (2005).
\newblock {\em Reassembling the {Social}: {An} {Introduction} to the {Actor}-{Network} {Theory}}.
\newblock Oxford University Press, Oxford, England and New York, NY, USA.

\bibitem[Lazar, 2024]{sobel_legitimacy_2024}
Lazar, S. (2024).
\newblock Legitimacy, {Authority}, and {Democratic} {Duties} of {Explanation}.
\newblock In Sobel, D. and Wall, S., editors, {\em Oxford {Studies} in {Political} {Philosophy} {Volume} 10}, pages 28--56. Oxford University PressOxford, 1 edition.

\bibitem[Lee, 1984]{lee_bureaucracies_1984}
Lee, R.~M. (1984).
\newblock Bureaucracies, bureaucrats and information technology.
\newblock {\em European Journal of Operational Research}, 18(3):293--303.

\bibitem[Lindblom, 1959]{lindblom_science_1959}
Lindblom, C.~E. (1959).
\newblock The {Science} of "{Muddling} {Through}".
\newblock {\em Public Administration Review}, 19(2):79--88.

\bibitem[Lipsky, 1980]{lipsky_street-level_1980}
Lipsky, M. (1980).
\newblock {\em Street-{Level} {Bureaucracy}: {The} {Dilemmas} of the {Individual} in {Public} {Service}}.
\newblock Russell Sage Foundation.

\bibitem[Lipton, 2017]{lipton_mythos_2017}
Lipton, Z.~C. (2017).
\newblock The {Mythos} of {Model} {Interpretability}.

\bibitem[Maslej et~al., 2024]{maslej_artificial_2024}
Maslej, N., Fattorini, L., Perrault, R., Parli, V., Reuel, A., Brynjolfsson, E., Etchemendy, J., Ligett, K., Lyons, T., Manyika, J., Niebles, J.~C., Shoham, Y., Wald, R., and Clark, J. (2024).
\newblock Artificial {Intelligence} {Index} {Report} 2024.

\bibitem[Matthias, 2004]{matthias_responsibility_2004}
Matthias, A. (2004).
\newblock The responsibility gap: {Ascribing} responsibility for the actions of learning automata.
\newblock {\em Ethics and Information Technology}, 6(3):175--183.

\bibitem[Mayerfeld, 1999]{mayerfeld1999suffering}
Mayerfeld, J. (1999).
\newblock {\em Suffering and moral responsibility}.
\newblock Oxford University Press.

\bibitem[Mellouli et~al., 2024]{mellouli_introduction_2024}
Mellouli, S., Janssen, M., and Ojo, A. (2024).
\newblock Introduction to the {Issue} on {Artificial} {Intelligence} in the {Public} {Sector}: {Risks} and {Benefits} of {AI} for {Governments}.
\newblock {\em Digit. Gov.: Res. Pract.}, 5(1):1:1--1:6.

\bibitem[Mokander and Schroeder, 2024]{mokander_2024_ArtificialIntelligenceRationalizationlimitscontrol}
Mokander, J. and Schroeder, R. (2024).
\newblock Artificial intelligence, rationalization, and the limits of control in the public sector: the case of tax policy optimization.
\newblock {\em Social Science Computer Review}, page 08944393241235175.

\bibitem[Muellerleile and Robertson, 2018]{muellerleile_digital_2018}
Muellerleile, C. and Robertson, S.~L. (2018).
\newblock Digital {Weberianism}: {Bureaucracy}, {Information}, and the {Techno}-rationality of {Neoliberal} {Capitalism}.
\newblock {\em Indiana Journal of Global Legal Studies}, 25(1):187--216.

\bibitem[Newman et~al., 2022]{newman_2022_DigitalTechnologiesArtificialintelligencebureaucratic}
Newman, J., Mintrom, M., and O'Neill, D. (2022).
\newblock Digital technologies, artificial intelligence, and bureaucratic transformation.
\newblock {\em Futures}, 136:102886.

\bibitem[Niskanen, 1971]{niskanen_bureaucracy_1971}
Niskanen, W.~A. (1971).
\newblock {\em Bureaucracy and {Representative} {Government}}.
\newblock Transaction Publishers.

\bibitem[Nissenbaum, 1996]{nissenbaum_accountability_1996}
Nissenbaum, H. (1996).
\newblock Accountability in a computerized society.
\newblock {\em Science and Engineering Ethics}, 2(1):25--42.

\bibitem[OECD, 2019]{OECDAI19}
OECD (2019).
\newblock Recommendation of the council on artificial intelligence.
\newblock Technical Report OECD/LEGAL/0449, Organisation for Economic Cooperation and Development (OECD) Legal Instruments, Paris.

\bibitem[Perry et~al., 2023]{Perry23}
Perry, N., Srivastava, M., Kumar, D., and Boneh, D. (2023).
\newblock Do users write more insecure code with {AI} assistants?
\newblock In {\em Proceedings of the 2023 ACM SIGSAC Conference on Computer and Communications Security}, CCS '23, page 2785–2799, New York, NY, USA. Association for Computing Machinery.

\bibitem[Popa, 2024]{popa_multilevel_2024}
Popa, D. (2024).
\newblock Multilevel oversight of {AI} systems in line with the {AI} {Act}.

\bibitem[Pozen, 2018]{pozen_transparencys_2018}
Pozen, D.~E. (2018).
\newblock Transparency's {Ideological} {Drift}.
\newblock {\em Yale Law Journal}.

\bibitem[Reuel et~al., 2024]{reuel_betterbench_2024}
Reuel, A., Hardy, A., Smith, C., Lamparth, M., Hardy, M., and Kochenderfer, M.~J. (2024).
\newblock {BetterBench}: {Assessing} {AI} {Benchmarks}, {Uncovering} {Issues}, and {Establishing} {Best} {Practices}.

\bibitem[Roehl and Crompvoets, 2025]{doi:10.1177/09520767231197801}
Roehl, U. and Crompvoets, J. (2025).
\newblock Inside algorithmic bureaucracy: {Disentangling} automated decision-making and good administration.
\newblock {\em Public Policy and Administration}, 40(2):322--350.

\bibitem[Santoni~de Sio and Van~den Hoven, 2018]{Santoni18}
Santoni~de Sio, F. and Van~den Hoven, J. (2018).
\newblock Meaningful human control over autonomous systems: A philosophical account.
\newblock {\em Frontiers in Robotics and AI}, 5.

\bibitem[Schmitz et~al., 2025]{schmitz-etal-2025-oversight}
Schmitz, C., Rystrøm, J., and Batzner, J. (2025).
\newblock Oversight structures for agentic {AI} in public-sector organizations.
\newblock In Kamalloo, E., Gontier, N., Lu, X.~H., Dziri, N., Murty, S., and Lacoste, A., editors, {\em Proceedings of the 1st workshop for research on agent language models ({REALM} 2025)}, pages 298--308, Vienna, Austria. Association for Computational Linguistics.

\bibitem[Scott, 1998]{scott_seeing_1998}
Scott, J.~C. (1998).
\newblock {\em Seeing {Like} a {State}: {How} {Certain} {Schemes} to {Improve} the {Human} {Condition} {Have} {Failed}}.
\newblock Yale University Press.

\bibitem[Sebo, 2018]{sebo_2018_MoralProblemOtherMinds}
Sebo, J. (2018).
\newblock The {Moral} {Problem} of {Other} {Minds}:.
\newblock {\em The Harvard Review of Philosophy}, 25:51--70.

\bibitem[Sebo and Long, 2023]{sebo_2023_MoralConsiderationAIsystems2030}
Sebo, J. and Long, R. (2023).
\newblock Moral consideration for {AI} systems by 2030.
\newblock {\em AI and Ethics}.

\bibitem[Senge, 1990]{senge_fifth_1990}
Senge, P.~M. (1990).
\newblock {\em The {Fifth} {Discipline}: {The} {Art} and {Practice} of the {Learning} {Organization}}.
\newblock Doubleday/Currency.

\bibitem[Shafir, 1985]{shafir_incongruity_1985}
Shafir, G. (1985).
\newblock The {Incongruity} between {Destiny} and {Merit}: {Max} {Weber} on {Meaningful} {Existence} and {Modernity}.
\newblock {\em The British Journal of Sociology}, 36(4):516--530.

\bibitem[Shank and DeSanti, 2018]{shank_2018_AttributionsMoralityMindartificialintelligence}
Shank, D.~B. and DeSanti, A. (2018).
\newblock Attributions of morality and mind to artificial intelligence after real-world moral violations.
\newblock {\em Computers in Human Behavior}, 86:401--411.

\bibitem[Simon, 1947]{simon_administrative_1947}
Simon, H.~A. (1947).
\newblock {\em Administrative {Behavior}}.
\newblock Macmillan Company.

\bibitem[Sparrow and Flenady, 2025]{sparrow_2025_TestimonyGapMachinesReasons}
Sparrow, R. and Flenady, G. (2025).
\newblock The {Testimony} {Gap}: {Machines} and {Reasons}.
\newblock {\em Minds and Machines}, 35(1):12.

\bibitem[Torrance, 2014]{torrance_2014_ArtificialConsciousnessArtificialEthicsRealism}
Torrance, S. (2014).
\newblock Artificial {Consciousness} and {Artificial} {Ethics}: {Between} {Realism} and {Social} {Relationism}.
\newblock {\em Philosophy \& Technology}, 27(1):9--29.

\bibitem[Troshani et~al., 2021]{Troshani03092021}
Troshani, I., Hill, S.~R., Sherman, C., and Arthur, D. (2021).
\newblock Do we trust in {AI}? {Role} of anthropomorphism and intelligence.
\newblock {\em Journal of Computer Information Systems}, 61(5):481--491.

\bibitem[UNESCO, 2022]{UNESCOairec22}
UNESCO (2022).
\newblock Recommendation on the ethics of artificial intelligence.
\newblock Technical Report SHS/BIO/PI/2021/1, United Nations Educational, Scientific and Cultural Organization (UNESCO), Paris.

\bibitem[Veluwenkamp and Hindriks, 2024]{Veluwenkamp03102024}
Veluwenkamp, H. and Hindriks, F. (2024).
\newblock Artificial agents: responsibility \& control gaps.
\newblock {\em Inquiry : a journal of medical care organization, provision and financing}, 0(0):1--25.

\bibitem[Verbeek, 2006]{verbeek_materializing_2006}
Verbeek, P.-P. (2006).
\newblock Materializing {Morality}: {Design} {Ethics} and {Technological} {Mediation}.
\newblock {\em Science, Technology, and Human Values}, 31(3):361--380.

\bibitem[Vredenburgh, 2023]{vredenburgh_ai_2023}
Vredenburgh, K. (2023).
\newblock {AI} and bureaucratic discretion.
\newblock {\em Inquiry}, 0(0):1--30.

\bibitem[Wallis, 2021]{Wallis21}
Wallis, N. (2021).
\newblock {\em The Great Post Office Scandal: The Fight to Expose A Multimillion Pound Scandal Which Put Innocent People in Jail}.
\newblock Bath Publishing Limited.

\bibitem[Weber, 1921]{weber1978economy}
Weber, M. (1921).
\newblock {\em Economy and society: {An} outline of interpretive sociology}.
\newblock Economy and society: {An} outline of interpretative sociology. University of California Press, 1978 edition, ed. roth, g \& wittich, c. edition.

\bibitem[Widder and Nafus, 2023]{widder_dislocated_2023}
Widder, D.~G. and Nafus, D. (2023).
\newblock Dislocated accountabilities in the “{AI} supply chain”: {Modularity} and developers’ notions of responsibility.
\newblock {\em Big Data \& Society}, 10(1):20539517231177620.

\bibitem[Winfield et~al., 2021]{winfield_ieee_2021}
Winfield, A. F.~T., Booth, S., Dennis, L.~A., Egawa, T., Hastie, H., Jacobs, N., Muttram, R.~I., Olszewska, J.~I., Rajabiyazdi, F., Theodorou, A., Underwood, M.~A., Wortham, R.~H., and Watson, E. (2021).
\newblock {IEEE} {P7001}: {A} {Proposed} {Standard} on {Transparency}.
\newblock {\em Frontiers in Robotics and AI}, 8.

\bibitem[Wirtz et~al., 2019]{wirtz_2019_ArtificialIntelligencePublicSectorApplications}
Wirtz, B.~W., Weyerer, J.~C., and Geyer, C. (2019).
\newblock Artificial {Intelligence} and the {Public} {Sector}—{Applications} and {Challenges}.
\newblock {\em International Journal of Public Administration}, 42(7):596--615.

\bibitem[Yam et~al., 2022]{yam_2022_WhenYourBossrobotWorkers}
Yam, K.~C., Goh, E.-Y., Fehr, R., Lee, R., Soh, H., and Gray, K. (2022).
\newblock When your boss is a robot: {Workers} are more spiteful to robot supervisors that seem more human.
\newblock {\em Journal of Experimental Social Psychology}, 102:104360.

\bibitem[Young et~al., 2021]{young_2021_ArtificialIntelligenceAdministrativeEvil}
Young, M.~M., Himmelreich, J., Bullock, J.~B., and Kim, K.-C. (2021).
\newblock Artificial {Intelligence} and {Administrative} {Evil}.
\newblock {\em Perspectives on Public Management and Governance}, 4(3):244--258.

\bibitem[Zacka, 2017]{zacka_2017_WhenStateMeetsStreetPublic}
Zacka, B. (2017).
\newblock {\em When the {State} {Meets} the {Street}: {Public} {Service} and {Moral} {Agency}}.
\newblock Harvard University Press.

\bibitem[Zacka, 2022]{zacka_2022_PoliticalTheoryRediscoversPublicAdministration}
Zacka, B. (2022).
\newblock Political {Theory} {Rediscovers} {Public} {Administration}.
\newblock {\em Annual Review of Political Science}, 25(Volume 25, 2022):21--42.

\bibitem[Zouridis et~al., 2019]{zouridis_automated_2019}
Zouridis, S., van Eck, M., and Bovens, M. (2019).
\newblock Automated {Discretion}.

\end{thebibliography}

\end{document}